\documentclass[pdflatex,sn-mathphys-num]{sn-jnl}

\usepackage{graphicx}%
\usepackage{multirow}%
\usepackage{amsmath,amssymb,amsfonts}%
\usepackage{amsthm}%
\usepackage{mathrsfs}%
\usepackage[title]{appendix}%
\usepackage[dvipsnames]{xcolor}%
\usepackage{textcomp}%
\usepackage{manyfoot}%
\usepackage{booktabs}%
\usepackage{algorithm}%
\usepackage{algorithmicx}%
\usepackage{algpseudocode}%
\usepackage{listings}%
\usepackage{refcount}
\usepackage{bm}
\usepackage{placeins}
\usepackage[normalem]{ulem}
\usepackage{cancel}
\usepackage{comment}
\usepackage{float}
\usepackage{centernot}
\usepackage{tikz}
\usetikzlibrary{backgrounds}
\usetikzlibrary{positioning,arrows.meta,matrix}
\usetikzlibrary{calc}
\usepackage{pdflscape}
\usepackage{eso-pic}

\newcommand{\teq}{\mathrel{\mkern-3mu=\mkern-3mu}}

\theoremstyle{thmstyleone}%
\newtheorem{theorem}{Theorem}
\newtheorem{proposition}[theorem]{Proposition}%
\newtheorem{lemma}[theorem]{Lemma}%

\theoremstyle{thmstyletwo}%
\newtheorem{remark}{Remark}%

\theoremstyle{thmstylethree}%
\newtheorem{definition}{Definition}%

\begin{document}
\title{Non-Uniqueness in the Coordinate-Based Normalization of Tetrahedral Similarity Classes}

\author[1]{\fnm{Miguel A.} \sur{Padrón}}
\email{miguel.padron@ulpgc.es}

\author[2]{\fnm{Agustín} \sur{Trujillo-Pino}}
\email{agustin.trujillo@ulpgc.es}

\author*[1]{\fnm{Jose P.} \sur{Suárez}}
\email{josepablo.suarez@ulpgc.es}

\affil[1]{%
\orgdiv{Department of Mathematics, Graphics and Computation; IUMA Institute},
\orgname{University of Las Palmas de Gran Canaria},
\orgaddress{%
\city{Las Palmas de Gran Canaria},
\postcode{35017},
\country{Spain}}}

\affil[2]{%
\orgdiv{Center of Image Technology; IUCES Institute},
\orgname{University of Las Palmas de Gran Canaria},
\orgaddress{%
\city{Las Palmas de Gran Canaria},
\postcode{35017},
\country{Spain}}}

\abstract{Recently, a new coordinate-based normalization has been introduced in \cite{michaud2026orbits} to represent tetrahedral similarity classes by normalized coordinates, enabling the study of LEB orbits as a discrete dynamical system. However, the uniqueness of this representation is not guaranteed when two or more edges adjacent to the selected longest edge have the same minimum length and remain admissible under the normalization criteria. In this paper, we demonstrate that different admissible choices may lead to distinct normalized coordinate representations of the same tetrahedral similarity class. We first present an explicit counterexample in which two admissible vertex labelings yield different normalized coordinate representations related by reflection, despite corresponding to the same similarity class. We then provide a systematic classification of the nine possible edge configurations arising in the selection of the canonical face when the longest edge of the tetrahedron is unique, and analyze the corresponding geometric and algebraic conditions. Our analysis identifies the configurations in which the normalization is non-unique and distinguishes them from those in which the ambiguity is resolved by the length criterion, angle criterion, or geometric symmetries. We further show that this non-uniqueness leads to an overcounting of distinct similarity classes.}



\maketitle

\maketitle

\section{Introduction}\label{sec1}

The Longest-Edge Bisection (LEB) algorithm is a widely used refinement technique for triangular and tetrahedral meshes \cite{Adl,KoPlSu,PaSuPla,RivLev, Sik}. In three dimensions, the algorithm subdivides a tetrahedron into two child tetrahedra by bisecting one of its longest edges. Despite the simplicity of the bisection procedure, the geometric analysis of the tetrahedra generated by successive  subdivisions remains a challenging problem, particularly with regard to the evolution of their geometric shapes and the number of distinct similarity classes produced during the subdivision process \cite{HaKoKri_2014,SuTrTa,TrSuPa}.

The
resulting sequence of similarity classes  will be referred to as the LEB orbit of the
initial tetrahedron. To facilitate their representation and computational analysis, a normalization procedure associates each tetrahedron with normalized coordinates in a prescribed configuration space. Thus, subdivision and normalization
play fundamentally different roles: subdivision determines the dynamics and hence the
descendants generated by LEB, whereas normalization provides a representation in
which tetrahedra belonging to the same similarity class can be compared and identified. 

The classification of tetrahedra into similarity classes plays an important role in the theoretical analysis of LEB \cite{Casado2015,HaKoKri_2014,TrSuPa}, since tetrahedra belonging to the same similarity class have identical geometric shapes up to rigid transformations, reflections, and uniform scaling. This classification allows the  subdivision process to be analyzed independently of the position, orientation, and size of each tetrahedron. In particular, the identification of distinct similarity classes is essential for investigating the structure of LEB orbits and the geometric behavior of tetrahedra under successive subdivisions \cite{adiprasito2026, PadronTrujilloSuarez2025}. The study of similarity classes is also of fundamental importance in the Finite Element Method. Limiting the number of distinct element shapes not only facilitates the reuse of element matrices and reduces computational costs, but also contributes to maintaining uniform geometric quality. Therefore, the reliability and robustness of the numerical approximation are enhanced.

Several approaches have been developed to represent and analyze tetrahedral similarity classes. One approach is based on the six squared edge lengths of a tetrahedron, which provide an algebraic description of its geometric shape \cite{PadronTrujilloSuarez2025, SuTrTa,TrSuPa}. An alternative approach is coordinate-based; see, for example, \cite{PerdomoPlaza2014}. Recently, Michaud and Korotov \cite{michaud2026orbits} proposed a coordinate-based normalization intended to represent tetrahedral similarity classes in normalized coordinates. Their construction provides a geometric framework for studying LEB orbits as a discrete dynamical system in a normalized-coordinate space.

The  normalization procedure begins by selecting a canonical face sharing one of the longest edges of the tetrahedron. Among the four edges adjacent to the selected longest edge,  an edge of minimum length is chosen to determine the canonical face. If the minimum-length condition is satisfied by more than one edge,  then the angle  criterion is applied to resolve the tie. When the angle criterion does not distinguish between the admissible choices, the selection is considered arbitrary. Once the canonical face has been determined, a sequence of translations, rotations, and uniform scalings transforms the tetrahedron into a normalized coordinate representation, thereby positioning the tetrahedron in a canonical orientation.

To provide a unique coordinate representation of each similarity class,  all admissible choices arising during the normalization procedure must lead to the same canonical representation. However, this property is not guaranteed when two or more edges adjacent to the selected longest edge have the same minimum length. In particular, an unresolved tie between admissible choices may produce different vertex labelings and, consequently, distinct normalized coordinate representations of the same tetrahedral similarity class.

The purpose of this paper is to investigate this lack of uniqueness and characterize the geometric configurations in which it arises. We first construct an explicit counterexample involving a tetrahedron whose canonical face is uniquely determined but is isosceles, so that either of its two edges of minimum length can be selected. In this case, selecting one edge or the other induces a relabeling of the vertices of the longest edge. We prove that these two choices lead to different triples, even though the corresponding tetrahedra are related by reflection and therefore belong to the same similarity class. This example establishes that the Michaud--Korotov normalization is not similarity-invariant in general and therefore does not necessarily assign a unique normalized representative to each tetrahedral similarity class.

We then extend this analysis to a systematic classification of the nine possible edge configurations. These arise from the number and combinatorial arrangement of the edges of minimum length adjacent to  the unique longest edge. For each configuration, we examine how the normalization criteria apply and determine whether the admissible choices yield a unique normalized representative or distinct representations of the same similarity class. 

The remainder of this paper is organized as follows. Section 2 presents the preliminaries and the sextuple-based representation of tetrahedral similarity classes. Section 3 describes the Michaud–Korotov normalization. Section 4 presents an explicit counterexample demonstrating the non-uniqueness of this normalization, and Section 5 classifies the possible edge configurations and determines when the normalized representation is unique. Section 6 discusses overcounting in LEB orbits. Finally, Section 7 presents the final remarks.

\section{Preliminaries}\label{sec1}

Let $\mathcal{T}$ denote the set of all non-degenerate tetrahedra in
$\mathbb{R}^{3}$. A tetrahedron $T\in\mathcal{T}$ is regarded as an unlabeled
geometric tetrahedron; that is, it is independent of the particular ordering of its
vertices.

\begin{definition}[Similarity of tetrahedra]
Let $T,T'\in\mathcal{T}$. We say that $T$ and $T'$ are \emph{similar},
and write $T\sim T', $
if there exist $\lambda>0$, $Q\in O(3)$, and $b\in\mathbb{R}^{3}$ such
that $ T'=\lambda Q(T)+b. $
Thus, two tetrahedra are similar if they differ by a uniform scaling,
an orthogonal transformation (including reflections), and a translation.
\end{definition}

The relation $\sim$ is an equivalence relation on $\mathcal{T}$. We
denote by $\mathcal{S}:=\mathcal{T}/\!\sim $
the corresponding quotient space of similarity classes, and by $  [T]\in\mathcal{S} $
the similarity class of $T\in\mathcal{T}$.

\begin{definition}[Longest Edge Bisection]
Let $T\in\mathcal{T}$ and let $E_{\max}(T)$ denote the set of edges of
$T$ having maximal length. For each edge  $e\in E_{\max}(T)$, let
$\mathcal{B}_{e}(T)=\{T_1,T_2\}$ denote the two tetrahedra obtained by
bisecting $e$ at its midpoint. The \emph{Longest Edge Bisection}
operator is the multivalued subdivision operator
\[
    \mathcal{B}_{\mathrm{LEB}}(T)
    =
    \bigcup_{e\in E_{\max}(T)}\mathcal{B}_{e}(T).
\]
When the longest edge is unique, this operator produces a uniquely
defined pair of child tetrahedra. If several edges have the same maximal
length, different admissible choices may lead to different subdivision patterns.
\end{definition}

In order to obtain a deterministic subdivision process in the presence
of ties, one may supplement the LEB rule with a deterministic
tie-breaking rule. We denote such a rule by 
\(\chi\),
where, for each \(T\in\mathcal T\), $\chi(T)\in E_{\max}(T). $
The corresponding deterministic LEB subdivision operator is then defined
by $\mathcal{B}_{\chi}(T) := \mathcal{B}_{\chi(T)}(T).$
Thus, the subdivision rule and the tie-breaking convention are
conceptually distinct: the former specifies that a longest edge must be
bisected, whereas the latter specifies which longest edge is selected
when the maximum is not unique. For example, in \cite{TrSuPa}, it was shown that the tie-breaking convention adopted by SCLEB, which is compatible with the criterion suggested by Adler \cite{Adl}, guarantees convergence of the
subdivision process for the nearly equilateral tetrahedra of the $R_1^+$ family.

Starting from $T_{0}\in\mathcal{T}$, the $n$-th generation of the
deterministic LEB subdivision is defined recursively by $\mathcal{B}_{\chi}^{\,0}(T_{0})=\{T_{0}\}, $
and
\[
    \mathcal{B}_{\chi}^{\,n+1}(T_{0})
    =
    \bigcup_{T\in\mathcal{B}_{\chi}^{\,n}(T_{0})}
    \mathcal{B}_{\chi}(T),
    \qquad n\geq0.
\]
For the multivalued formulation, the same definition is obtained by
replacing $\mathcal{B}_{\chi}$ with $\mathcal{B}_{\mathrm{LEB}}$.

\begin{definition}[LEB orbit]
Assume that \(\chi\) is similarity-compatible, so that successive deterministic subdivisions of similar tetrahedra generate the same similarity classes.
Then the \emph{LEB orbit} of the similarity class $[T_{0}]$ is the set
\[
    \mathcal{O}_{\mathrm{LEB}}([T_{0}])
    :=
    \bigcup_{n\geq0}
    \left\{
        [T]\;:\;
        T\in\mathcal{B}_{\chi}^{\,n}(T_{0})
    \right\}.
\]
Thus, the LEB orbit records the distinct similarity classes generated
by successive applications of the subdivision rule, independently of
translation, rotation, reflection, scaling, and vertex labeling.
\end{definition}

For computational purposes, it is convenient to represent similarity classes by elements of a normalized configuration space. Let
$\mathcal{X}$ be such a space and let $  \mathcal{N}:\mathcal{T}\longrightarrow\mathcal{X}$ be a \emph{normalization map}. Its purpose is to assign to each tetrahedron a representative in $\mathcal{X}$. 

\begin{definition}[Similarity-invariant normalization]
\label{def:similarity}
A normalization map
$\mathcal{N}:\mathcal{T}\rightarrow\mathcal{X}$ is
\emph{similarity-invariant} if $
    T\sim T'
    \quad\Longrightarrow\quad
    \mathcal{N}(T)=\mathcal{N}(T').
$
In this case, $\mathcal{N}$ induces a well-defined map $
    \widehat{\mathcal{N}}:\mathcal{S}\longrightarrow\mathcal{X},  \quad
    \widehat{\mathcal{N}}([T])=\mathcal{N}(T). $
\end{definition}

We will show in Section 4 that the coordinate-based normalization procedure considered in~\cite{michaud2026orbits} does not satisfy Definition 4 in general, and therefore does not define a canonical normalization in the sense of Definition 5.

\begin{definition}[Canonical normalization]
\label{def:canonical}
A similarity-invariant normalization
$\mathcal{N}:\mathcal{T}\rightarrow\mathcal{X}$ is said to be
\emph{canonical} if $
    \mathcal{N}(T)=\mathcal{N}(T')
    \quad\Longrightarrow\quad
    T\sim T'.
$
Equivalently, the induced map
$
    \widehat{\mathcal{N}}:\mathcal{S}\longrightarrow\mathcal{X}
$
is injective. Hence, a canonical normalization assigns a unique
normalized representative to each similarity class, with distinct similarity classes having distinct normalized representatives.
\label{canonical}
\end{definition}

Similarity invariance ensures that  similar tetrahedra receive the same normalized representation, whereas injectivity ensures that non-similar tetrahedra receive different representations. Hence, a canonical normalization establishes a one-to-one correspondence between tetrahedral similarity classes and normalized representatives in the image of \(\widehat{\mathcal N}\).

The distinction between subdivision, similarity classes, and normalized
representatives is essential. The LEB operator determines the geometric
dynamics, the quotient map identifies similar
tetrahedra, and the normalization provides a computational
representation of the resulting similarity classes.

\begin{remark}
Let $\pi:\mathcal{T}\longrightarrow\mathcal{S}$,\quad 
$\pi(T)=[T]$, denote the natural quotient map.  
The subdivision process can then be
schematically represented as
$
    \mathcal{T}
    \xrightarrow{\;\mathcal{B}_{\chi}\;}
    \mathcal{T}
    \xrightarrow{\;\pi\;}
    \mathcal{S}
    \xrightarrow{\;\widehat{\mathcal{N}}\;}
    \mathcal{X}.
$

\end{remark}

This decomposition separates the generation of tetrahedra from their
geometric identification and from their computational representation.

\begin{definition}[Normalized LEB orbit]
Given $T_{0}\in\mathcal{T}$, the \emph{normalized LEB orbit} is
\[
    \mathcal{O}_{\mathcal{N}}(T_{0})
    :=
    \bigcup_{n\geq0}
    \left\{
        \mathcal{N}(T)\;:\;
        T\in\mathcal{B}_{\chi}^{\,n}(T_{0})
    \right\}.
\]
\end{definition}

If $\widehat{\mathcal N}$ is well defined and injective on
$\mathcal O_{\mathrm{LEB}}([T_0])$, then the normalized orbit and the orbit
in the quotient space have the same cardinality:
\[
\left|
\mathcal O_{\mathcal N}(T_0)
\right|
=
\left|
\mathcal O_{\mathrm{LEB}}([T_0])
\right|.
\]

If \(\mathcal N\) is not similarity-invariant, a single similarity class may have several normalized
representatives, and consequently, in some cases,
$
\left|
\mathcal{O}_{\mathcal{N}}(T_{0})
\right|
>
\left|
\mathcal{O}_{\mathrm{LEB}}([T_{0}])
\right|
$. 

On the other hand, if \(\widehat{\mathcal N}\) is well defined but not injective, distinct similarity classes may have the same normalized representative, and consequently
$
\left|
\mathcal{O}_{\mathcal{N}}(T_{0})
\right|
<
\left|
\mathcal{O}_{\mathrm{LEB}}([T_{0}])
\right|
$
may occur.

Therefore, the number of normalized states obtained in a
computational experiment cannot, by itself, be interpreted as the
number of distinct similarity classes unless the 
induced map \(\widehat{\mathcal N}\) is known to be well defined and injective on the generated similarity classes.

\begin{proposition}[Canonical normalization and orbit counting]
Let $\mathcal{N}:\mathcal{T}\rightarrow\mathcal{X}$ be a
similarity-invariant normalization and let
$T_{0}\in\mathcal{T}$. If the induced map
$
    \widehat{\mathcal{N}}:\mathcal{S}\rightarrow\mathcal{X}
$
is injective on $\mathcal{O}_{\mathrm{LEB}}([T_{0}])$, then the map
$
    [T]\longmapsto\mathcal{N}(T)
$
defines a bijection between
$\mathcal{O}_{\mathrm{LEB}}([T_{0}])$ and
$\mathcal{O}_{\mathcal{N}}(T_{0})$. In particular,
$
    \left|
        \mathcal{O}_{\mathrm{LEB}}([T_{0}])
    \right|
    =
    \left|
        \mathcal{O}_{\mathcal{N}}(T_{0})
    \right|.
$
\end{proposition}

\begin{proof}
Similarity invariance ensures that $\mathcal{N}(T)$ depends only on
the similarity class $[T]$, so that $\widehat{\mathcal{N}}$ is
well-defined. Injectivity on
$\mathcal{O}_{\mathrm{LEB}}([T_{0}])$ implies that two distinct
similarity classes occurring in the orbit cannot have the same
normalized representative. By definition of
$\mathcal{O}_{\mathcal{N}}(T_{0})$, every normalized representative
corresponds to a similarity class in
$\mathcal{O}_{\mathrm{LEB}}([T_{0}])$. Hence, the induced map is a
bijection between the two orbits, and 
therefore they have the same cardinality.
\end{proof}






\subsection{Sextuple-based representation}

Let $T$ be a tetrahedron represented by the sextuple $s(T)=(A,B,C,D,E,F)$,
where \(A,\ldots,F\) denote the squared lengths of its six edges. The edges \(A,B,C\) form one face, and the edges opposite to \(A,B,C\) are \(F,D,E\), respectively.

For brevity, throughout the paper, parentheses will denote a particular sextuple representation of a tetrahedron, so that $T=(A,B,C,D,E,F)$, whereas brackets will denote its similarity class. Thus,
$[T]=[A,B,C,D,E,F]$
denotes a similarity class, which is invariant under translation, rotation, reflection and scaling.

\begin{definition}(Canonical sextuple representation). 
\label{def:sextuple} 
Let \(s(T)=(A,B,C,D,E,F)\) be the sextuple representation of the tetrahedron $T$. A canonical sextuple representative of its similarity class is obtained by fixing both an edge ordering and a scale as follows:

1. \emph{Edge ordering}. Among all admissible reorderings of the six edge values, select the sextuple for which the first component  corresponds to the longest edge, and the second component corresponds to the longest edge adjacent to it. Thus, the ordered sextuple 
always satisfies $A\geq B,C,D,E,F$ and $B\geq C,D,E$. 
If several admissible reorderings satisfy these conditions, choose the lexicographically largest one. 
Denoting \(\mathcal P(s(T))\) the set of admissible reorderings of \(s(T)\), the ordered representative is
$(A,B,C,D,E,F)=\max_{\mathrm{lex}} \mathcal P(s(T))$.

2. \emph{Scaling}. All sextuples of the form
$k(A,B,C,D,E,F)$, with $k>0$,
represent the same similarity class. To obtain a unique representative, two options are possible: the first one is to divide all components by \(A\), so that the first component is always equal to \(1\) and the remaining components do not exceed \(1\).
The second one applies when all the squared edge lengths are proportional to integers. In this case, an alternative exact representation is obtained by choosing the primitive integer sextuple, obtained by dividing all integer components by their greatest common divisor.



\end{definition}

Note that the canonical sextuple representation in Definition \ref{def:sextuple} satisfies the requirements of Definitions \ref{def:similarity} and \ref{def:canonical}. Any two similar tetrahedra yield the same canonical sextuple, independently of their position, orientation, reflection, initial edge ordering, or uniform scale. Conversely, two tetrahedra with the same canonical sextuple have identical edge-length ratios and therefore belong to the same similarity class. Hence, the sextuple representation establishes a one-to-one correspondence between tetrahedral similarity classes and canonical sextuples.


 


Using this sextuple representation, each successive LEB subdivision can be computed directly. For a class $[T]=[A,B,C,D,E,F]$,
the two children generated by bisecting the longest edge are represented by

$$[T_1]= [A,\;4B,\;2B+2C-A,\;2E+2D-A,\;4E,\;4F], $$
$$[T_2]= [A,\;2B+2C-A,\;4C,\;4D,\;2E+2D-A,\;4F]. $$

The Similarity Classes Longest-Edge Bisection (SCLEB) procedure \cite{TrSuPa}  consists of computing these two sextuples and then reordering them as defined above. Similarity classes can be distinguished by comparing their sextuples. 

When the squared edge lengths are integers, successive LEB subdivisions can be performed entirely with integer arithmetic. This makes the sextuple-based procedure substantially faster than the coordinate-based approach, while also avoiding numerical errors in the identification of similarity classes.

\section{The Michaud–Korotov normalization}

The normalization in \cite{michaud2026orbits} is based on selecting a canonical face of the tetrahedron and subsequently applying translations, orthogonal transformations, and uniform scaling to obtain a normalized configuration of its vertices.
We recall below the main steps of the procedure.

\subsection{Selection of the canonical face}

Let \(T\) be a tetrahedron with vertices \(V_1,V_2,V_3,V_4\). 
 The edge 
$V_1V_2$ is selected among the longest edges of the tetrahedron.
Then, $V_3$ is chosen so that 
$V_1V_3$ is the shortest edge among the four edges adjacent to 
$V_1V_2$. The triangle 
$V_1V_2V_3$ defines the canonical face, and the remaining vertex is denoted by \(V_4\).

When several adjacent edges have the same minimum length, more than one edge satisfies the length criterion. In such situations, an additional angle criterion is applied to distinguish between the admissible choices. If the  angle criterion also fails to resolve the tie, the selection is considered arbitrary.

An important observation is that the ambiguity may concern not only the selection of the canonical face itself, but also the labeling of its vertices.
In particular, if the canonical face is an isosceles triangle satisfying $
|V_1V_3|=|V_2V_3|$, 
interchanging the labels \(V_1\) and \(V_2\)  leaves the selected triangular face unchanged but produces different normalized coordinates for $V_4$.
This situation will be examined explicitly in Section \ref{sec:counterexample-uniqueness}.

\subsection{Construction of the normalized coordinate representation}


Once the canonical face 
\(V_1V_2V_3\) has been selected, the tetrahedron is translated, reoriented, and uniformly scaled so that the edge \(V_1V_2\) becomes the unit segment on the positive real axis. Vertex \(V_3\) lies in the upper half of the complex plane while $V_4$ lies in the upper half-space $z>0$.

After translating the tetrahedron so that  \(V_1\) coincides with the origin, we obtain
\[
\mathbf{b}'=V_2-V_1,
\qquad
\mathbf{c}'=V_3-V_1
\quad
\text{and}
\quad
\mathbf{d}'=V_4-V_1.
\]

Note that \(\mathbf b'\) corresponds to the  longest edge \(V_1V_2\), whereas \(\mathbf c'\) represents the edge of minimum length \(V_1V_3\).

An orthonormal basis is then constructed from the canonical face. 
The first basis vector is aligned with the edge \(V_1V_2\) as follows, 
$
\mathbf{e}_1=
\frac{\mathbf{b}'}{\|\mathbf{b}'\|}.
$
The second vector is obtained by subtracting from \(\mathbf c'\) its projection onto \(\mathbf e_1\):
\[
  \mathbf{v}_2=\mathbf{c}'-
(\mathbf{c}'\cdot\mathbf{e}_1)\mathbf{e}_1,\,\rightarrow \mathbf{e}_2=
\frac{\mathbf{v}_2}{\|\mathbf{v}_2\|}.
\]
Finally, the third basis vector is defined as the cross product 
\(
\mathbf{e}_3=
\mathbf{e}_1\times\mathbf{e}_2.
\)

The corresponding orthogonal transformation is 
given by the matrix
\(
R=
\begin{pmatrix}
\mathbf{e}_1^{T}\\
\mathbf{e}_2^{T}\\
\mathbf{e}_3^{T}
\end{pmatrix}
\)
yielding $\mathbf{b}''=R\mathbf{b}'$,
$\mathbf{c}''=R\mathbf{c}'$, and $\mathbf{d}''=R\mathbf{d}'$.
Dividing by the length of the longest edge gives the normalized coordinates
$\mathbf{b}'''=
\frac{\mathbf{b}''}{\|\mathbf{b}'\|}$,
$ 
\mathbf{c}'''=
\frac{\mathbf{c}''}{\|\mathbf{b}'\|}, 
\;\text{ and }\;
\mathbf{d}'''=
\frac{\mathbf{d}''}{\|\mathbf{b}'\|}.
$
Thus, the resulting tetrahedron in canonical position is represented by the new normalized vertices
\[
 P_1=(0,0,0),\quad
 P_2=(1,0,0),\quad
P_3=(c'''_1,c'''_2,0)\quad\text{and}\quad
P_4=(d'''_1,d'''_2,d'''_3).
\]

Using the complex-coordinate notation introduced by Michaud and Korotov, we define 
$
z=c'''_1+c'''_2i,
\;
w=d'''_1+d'''_2i
\; \text{ and } \;
t=|d'''_3|$.
The absolute value in the definition of \(t\) places $P_4$ in the upper half-space, independently of the orientation of the original tetrahedron.
The normalized representation is therefore expressed as the triple $
  \tau=(z,w,t).
$

Equivalently,  this representation can also be determined by the normalized quintuple
$
\left(
\operatorname{Re}(z),
\operatorname{Im}(z),
\operatorname{Re}(w),
\operatorname{Im}(w),
t
\right).
$
The first two vertices $P_1$ and $P_2$  remain fixed, while the remaining vertices $P_3$ and $P_4$ encode the geometry of the tetrahedron. 

\section{An explicit counterexample to uniqueness}\label{sec:counterexample-uniqueness}

Consider the tetrahedron
$T=(8,7,6,5,5,4)$, which can be represented by
\[
V_1=(0,0,0),\,\,
V_2=(\sqrt{8},0,0),\,\,
V_3=(\sqrt{2},\sqrt{3},0)\,\,\text{and}\,\,
V_4=
\left(
\frac{9\sqrt{2}}{8},
\frac{7\sqrt{3}}{12},
\frac{\sqrt{1986}}{24}
\right).
\]

Indeed, the squared edge lengths are
\[
\begin{gathered}
\|V_1V_2\|^2=8,\quad
\|V_2V_3\|^2=5,\quad
\|V_1V_3\|^2=5,\\
\|V_1V_4\|^2=7,\quad
\|V_2V_4\|^2=6\quad\text{and}\quad
\|V_3V_4\|^2=4.
\end{gathered}
\]

Thus, 
$V_1V_2$ is the unique longest edge, whereas the two edges of minimum length adjacent to it are $V_1V_3$ and $V_2V_3$,
both having squared length \(5\). Therefore, the canonical face is the triangle
 $V_1V_2V_3$,  but its two edges of minimum length give rise to two admissible canonical labelings of that face.

 We now apply the Michaud–Korotov normalization procedure to the two admissible labelings of the canonical face, arising from the choice of either of its two edges of minimum length.

\textbf{First choice:}  $\mathbf{V_1V_3}$. 

\begin{figure}[!hb]
  \centering
  \includegraphics[width=1.0\linewidth]{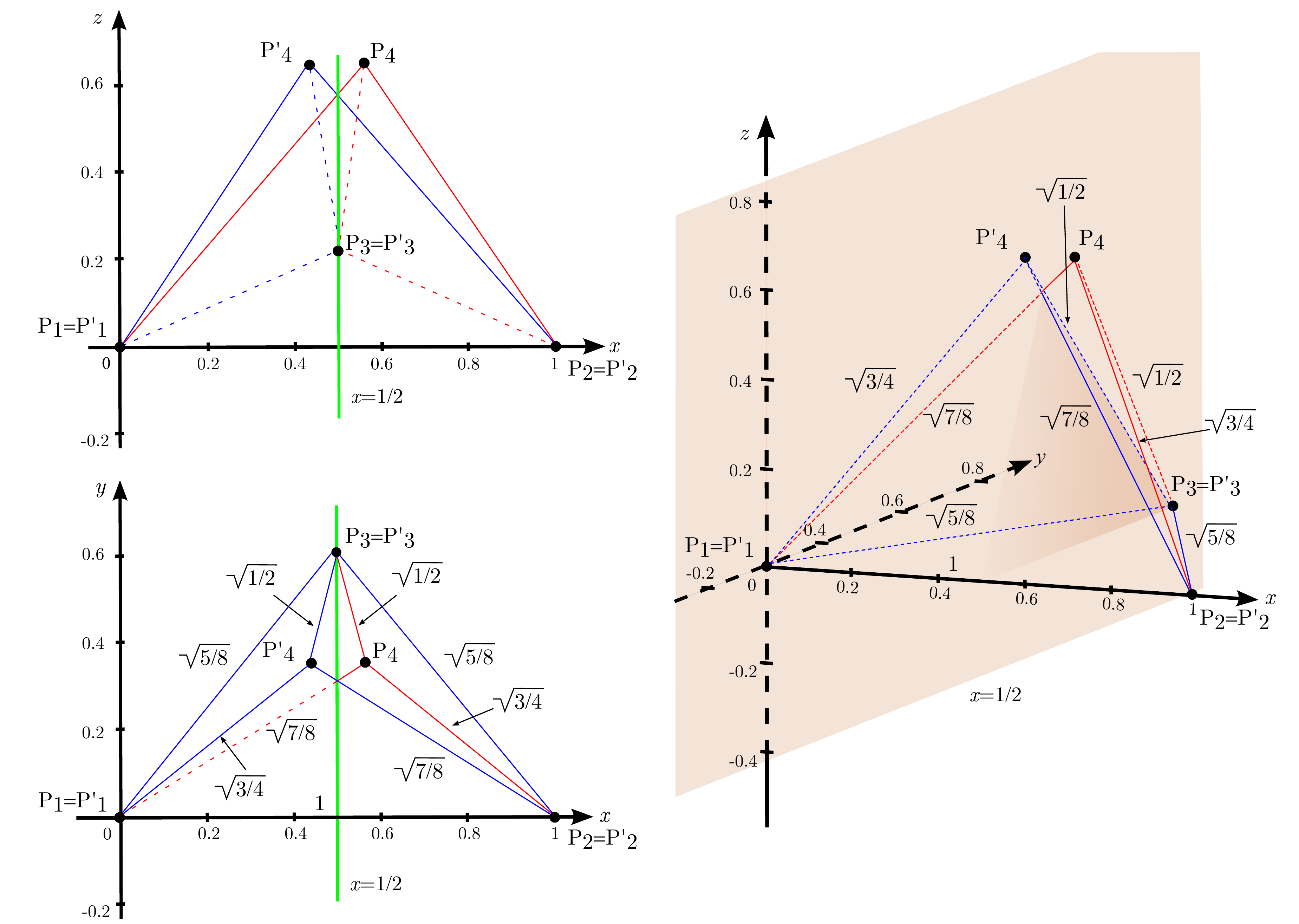}
  \caption{Tetrahedron $T$ and its reflected image in the normalized coordinate representation, shown from  different views. The reflection is across  the plane \(x=1/2\). 
  Both tetrahedra belong to the same similarity class.}
  \label{fig:fallo-korotov}
\end{figure}

\begin{enumerate}
  \item \textit{Translation of \(V_1\) to the origin:}
\[
\mathbf{b}'=V_2-V_1=
\begin{pmatrix}
\sqrt{8}\\
0\\
0
\end{pmatrix},
\,\,
\mathbf{c}'=V_3-V_1=
\begin{pmatrix}
\sqrt{2}\\
\sqrt{3}\\
0
\end{pmatrix},
\,\,
\text{and}
\,\,
\mathbf{d}'=V_4-V_1=
\begin{pmatrix}
\dfrac{9\sqrt{2}}{8}\\[2mm]
\dfrac{7\sqrt{3}}{12}\\[2.4mm]
\dfrac{\sqrt{1986}}{24}
\end{pmatrix}
\]

\item \textit{Orthonormal basis and normalization:}

Note that $T$ is already in the canonical orientation, since \(V_1\) is at the origin, \(V_1V_2\) is aligned with the positive \(x\)-axis, and the selected edge \(V_1V_3\) lies in the first quadrant of the plane \(z=0\). Hence, the orthonormal basis coincides with the canonical basis, so \(R=I\). Thus, \(\mathbf b''=\mathbf b'\), \(\mathbf c''=\mathbf c'\), and \(\mathbf d''=\mathbf d'\). After scaling by \(\|\mathbf b'\|=\sqrt8\), we obtain
\[
\mathbf{b}'''
=
\frac{\mathbf{b}''}{\|\mathbf{b}'\|}
=
\begin{pmatrix}
1\\
0\\
0
\end{pmatrix},\;\;
\mathbf{c}'''
=
\frac{\mathbf{c}''}{\|\mathbf{b}'\|}
=
\begin{pmatrix}
\dfrac12\\[2.4mm]
\dfrac{\sqrt6}{4}\\[2mm]
0
\end{pmatrix},\,\, \text{and}\,\,
\mathbf{d}'''
=
\frac{\mathbf{d}''}{\|\mathbf{b}'\|}
=
\begin{pmatrix}
\dfrac{9}{16}\\[2.4mm]
\dfrac{7\sqrt6}{48}\\[2.4mm]
\dfrac{\sqrt{993}}{48}
\end{pmatrix}.
\]
Finally,
\[
z=c_1'''+ic_2'''
=
\frac12+\frac{\sqrt6}{4}i,\quad w=d_1'''+id_2'''
=
\frac{9}{16}+\frac{7\sqrt6}{48}i,\;\text{ and }\; t=|d_3'''|
=
\frac{\sqrt{993}}{48}.
\]
Hence,
\begin{equation}~\label{eq:tau}
\tau=
\left(
\frac12+\frac{\sqrt6}{4}i,\left(
\frac{9}{16}+\frac{7\sqrt6}{48}i,
\frac{\sqrt{993}}{48}\right)
\right).
\end{equation}
The corresponding normalized vertices are $P_1=(0,0,0)$, $P_2=(1,0,0)$, $P_3=\left(\displaystyle\frac12,\frac{\sqrt6}{4},0\right)$, $P_4=\left(\displaystyle\frac{9}{16},\frac{7\sqrt{6}}{48},\frac{\sqrt{993}}{48}\right)$, as shown in Figure~\ref{fig:fallo-korotov}.
\end{enumerate}

\textbf{Second choice: $\mathbf{V_2V_3}$.}  

\begin{enumerate}
\item \textit{Translation of $V_2$ to the origin:}
\[
\mathbf{b}'
=V_1-V_2=
\begin{pmatrix}
-\sqrt{8}\\
\phantom{-}0\\
\phantom{-}0
\end{pmatrix},\,\,
\mathbf{c}'
=V_3-V_2=\begin{pmatrix}
-\sqrt{2}\\
\phantom{-}\sqrt{3}\\
\phantom{-}0
\end{pmatrix}\,\,\text{and}\,\,
\mathbf{d}'
=V_4-V_2=\begin{pmatrix}
-\dfrac{7\sqrt2}{8}\\[2mm]
\phantom{-}\dfrac{7\sqrt3}{12}\\[2.4mm]
\phantom{-}\dfrac{\sqrt{1986}}{24}
\end{pmatrix}.
\]
Unlike the first choice, in this case the tetrahedron \(T\) is not initially in the canonical orientation, 
and the vertices $V_1$ and $V_2$ are relabeled.
Therefore, we construct the orthonormal basis and the corresponding rotation matrix \(R\).

\item \textit{Orthonormal basis and normalization:}
\[
\mathbf{e}_1
=
\frac{\mathbf{b}'}{\|\mathbf{b}'\|}
=
\begin{pmatrix}
-1\\
\phantom{-}0\\
\phantom{-}0
\end{pmatrix},
\;\;\mathbf{v}_2
=
\mathbf{c}'-(\mathbf{c}'\cdot\mathbf{e}_1)\mathbf{e}_1
=
\begin{pmatrix}
-\sqrt2\\
\phantom{-}\sqrt3\\
\phantom{-}0
\end{pmatrix}
-
\sqrt2
\begin{pmatrix}
-1\\
\phantom{-}0\\
\phantom{-}0
\end{pmatrix}
=
\begin{pmatrix}
0\\
\sqrt3\\
0
\end{pmatrix}
\]
\[
\mathbf{e}_2
=
\frac{\mathbf{v}_2}{\|\mathbf{v}_2\|}
=
\begin{pmatrix}
0\\
1\\
0
\end{pmatrix},\,\,\text{and}\,\,
\mathbf{e}_3
=
\mathbf{e}_1\times\mathbf{e}_2
=
\begin{pmatrix}
-1\\
\phantom{-}0\\
\phantom{-}0
\end{pmatrix}
\times
\begin{pmatrix}
0\\
1\\
0
\end{pmatrix}
=
\begin{pmatrix}
\phantom{-}0\\
\phantom{-}0\\
-1
\end{pmatrix}
\]
The rotation matrix \(R\) is obtained from the new orthonormal basis, $R=
\begin{pmatrix}
\mathbf{e}_1^{T}\\
\mathbf{e}_2^{T}\\
\mathbf{e}_3^{T}
\end{pmatrix}$.
Then,
\[
\mathbf{b}''
=  
R\mathbf{b}'
=
\begin{pmatrix}
-1&\phantom{-}0&\phantom{-}0\\
\phantom{-}0&\phantom{-}1&\phantom{-}0\\
\phantom{-}0&\phantom{-}0&-1
\end{pmatrix}
\begin{pmatrix}
-\sqrt8\\
\phantom{-}0\\
\phantom{-}0
\end{pmatrix} 
= 
\begin{pmatrix}
\sqrt8\\
\phantom{-}0\\
\phantom{-}0
\end{pmatrix},
\]
\[
\mathbf{c}''
=
R\mathbf{c}'
=
\begin{pmatrix}
-1&\phantom{-}0&\phantom{-}0\\
\phantom{-}0&\phantom{-}1&\phantom{-}0\\
\phantom{-}0&\phantom{-}0&-1
\end{pmatrix}
\begin{pmatrix}
-\sqrt2\\
\phantom{-}\sqrt3\\
\phantom{-}0
\end{pmatrix}
=
\begin{pmatrix}
\sqrt2\\
\sqrt3\\
0
\end{pmatrix},\,\,\text{and}\,\,
\]
\[
\mathbf{d}''
=
R\mathbf{d}'
=
\begin{pmatrix}
-1&\phantom{-}0&\phantom{-}0\\
\phantom{-}0&\phantom{-}1&\phantom{-}0\\
\phantom{-}0&\phantom{-}0&-1
\end{pmatrix}
\begin{pmatrix}
-\dfrac{7\sqrt2}{8}\\[2mm]
\phantom{-}\dfrac{7\sqrt3}{12}\\[2.4mm]
\phantom{-}\dfrac{\sqrt{1986}}{24}
\end{pmatrix}
=
\begin{pmatrix}
\phantom{-}\dfrac{7\sqrt2}{8}\\[2mm]
\phantom{-}\dfrac{7\sqrt3}{12}\\[2.4mm]
-\dfrac{\sqrt{1986}}{24}
\end{pmatrix}.
\]
After scaling by \(\|\mathbf b'\|=\sqrt8\), we obtain
\[
  \mathbf{b}'''
=
\frac{\mathbf{b}''}{\|\mathbf{b}'\|}
=
\begin{pmatrix}
1\\
0\\
0
\end{pmatrix},\;\;
\mathbf{c}'''
=
\frac{\mathbf{c}''}{\|\mathbf{b}'\|}
=
\begin{pmatrix}
\dfrac12\\[2mm]
\dfrac{\sqrt6}{4}\\[2mm]
0
\end{pmatrix},\,\,\text{and}\,\,
\mathbf{d}'''
=
\frac{\mathbf{d}''}{\|\mathbf{b}'\|}
=
\begin{pmatrix}
\phantom{-}\dfrac{7}{16}\\[2mm]
\phantom{-}\dfrac{7\sqrt6}{48}\\[2.4mm]
-\dfrac{\sqrt{993}}{48}
\end{pmatrix}.
\]
Finally,
\[
z'
=
c_1'''+ic_2'''
=\frac{1}{2}+\frac{\sqrt{6}}{4}i,\quad
w'
=
d_1'''+id_2'''
=
\frac{7}{16}+\frac{7\sqrt6}{48}i,\;\text{ and }\;
t'
=|d'''_3|
=
\frac{\sqrt{993}}{48}.
\]
 Therefore,
\begin{equation}~\label{eq:tau'}
\tau'=
\left(
\frac12+\frac{\sqrt6}{4}i,
\left(\frac{7}{16}+\frac{7\sqrt6}{48}i,
\frac{\sqrt{993}}{48}\right)
\right).
\end{equation}
The corresponding normalized vertices are $P'_1=(0,0,0)$, $P'_2=(1,0,0)$,$P'_3=\left(\displaystyle\frac12,\frac{\sqrt6}{4},0\right)$, $P'_4=\left(\displaystyle\frac{7}{16},\frac{7\sqrt{6}}{48},\frac{\sqrt{993}}{48}\right)$, as shown in Figure~\ref{fig:fallo-korotov}.
\end{enumerate}

Clearly $\tau\neq\tau'$,
since the real part of the second complex coordinate is different.

\begin{proposition}
The Michaud--Korotov normalization is not similarity-invariant in general and therefore does not define a canonical normalization in the sense of Definition~5. 
\end{proposition}

\begin{proof}
The similarity class $[8,7,6,5,5,4]$ provides a counterexample. Indeed, there exist $T \sim T'$
such that
$
\mathcal N(T)=\tau$ given by (\ref{eq:tau}), and
$\mathcal N(T')=\tau'$ given by (\ref{eq:tau'}),
with
$\tau\neq\tau'$.
Hence,
$T\sim T'$ but
$\mathcal N(T)\neq\mathcal N(T'),$
so the normalization fails to satisfy the similarity-invariance condition of Definition~4.

\end{proof}

Thus, equality of normalized coordinate representations implies
$
\mathcal N(T)=\mathcal N(T')
\Longrightarrow
T\sim T'$,
but the converse implication does not hold in general. In particular,
$
T\sim T'\centernot\Longrightarrow
\mathcal N(T)=\mathcal N(T').
$

Therefore, the Michaud--Korotov normalization is not similarity-invariant in general and hence does not induce a well-defined map on similarity classes. Consequently, it is not canonical in the sense of Definition~5. Moreover, as discussed after Definition~6, this lack of similarity invariance may lead to several normalized coordinate representations corresponding to the same similarity class, and hence to an overcounting of normalized states.


Figure~\ref{fig:fallo-korotov} shows the tetrahedron $T$ in the normalized coordinate representation $\tau$, whose vertices are $P_1,P_2,P_3,P_4$, and its reflected image 
with respect to the plane \(x=1/2\), in the other normalized coordinate representation  $\tau'$, whose vertices are $P'_1,P'_2,P'_3,P'_4$. This reflection  is given by $\mathcal R(x,y,z)=(1-x,y,z)$  and maps these vertices as follows: 
\[
\mathcal R(0,0,0)=(1,0,0),\quad
\mathcal R(1,0,0)=(0,0,0),\quad
\mathcal R\!\left(\frac12,\frac{\sqrt6}{4},0\right)
=
\left(\frac12,\frac{\sqrt6}{4},0\right),\;{\text{and}}
\]
\[
\mathcal R\!\left(
\frac9{16},
\frac{7\sqrt6}{48},
\frac{\sqrt{993}}{48}
\right)
=
\left(
\frac7{16},
\frac{7\sqrt6}{48},
\frac{\sqrt{993}}{48}
\right).
\]
So, $\mathcal R(P_1)=P'_2$, $\mathcal R(P_2)=P'_1$, $\mathcal R(P_3)=P'_3$, and $\mathcal R(P_4)=P'_4$. Note that $P_1=P'_1$, $P_2=P'_2$, and $P_3=P'_3$. Nevertheless,   only $P_3$ remains invariant under the reflection, while $P_1$ and $P_2$ are exchanged and $P_4$ is mapped onto $P'_4$. Since $\mathcal R^2=I$, the reverse correspondence also holds.

The same reflection can also be expressed in the normalized coordinate representation by \(\mathcal R_\tau(z,w,t)=\left(1-\overline z,1-\overline w,t\right)\).  
Thus, \(\mathcal R\) and \(\mathcal R_\tau\) describe the same reflection in their respective representations. Note that  in this counterexample \(\mathcal R_\tau(\tau)=\tau'\). 

\section{Classification of the possible edge configurations}
\label{sec:classification}

\begin{figure}[htbp]
  \centering
  \includegraphics[width=0.83\linewidth]{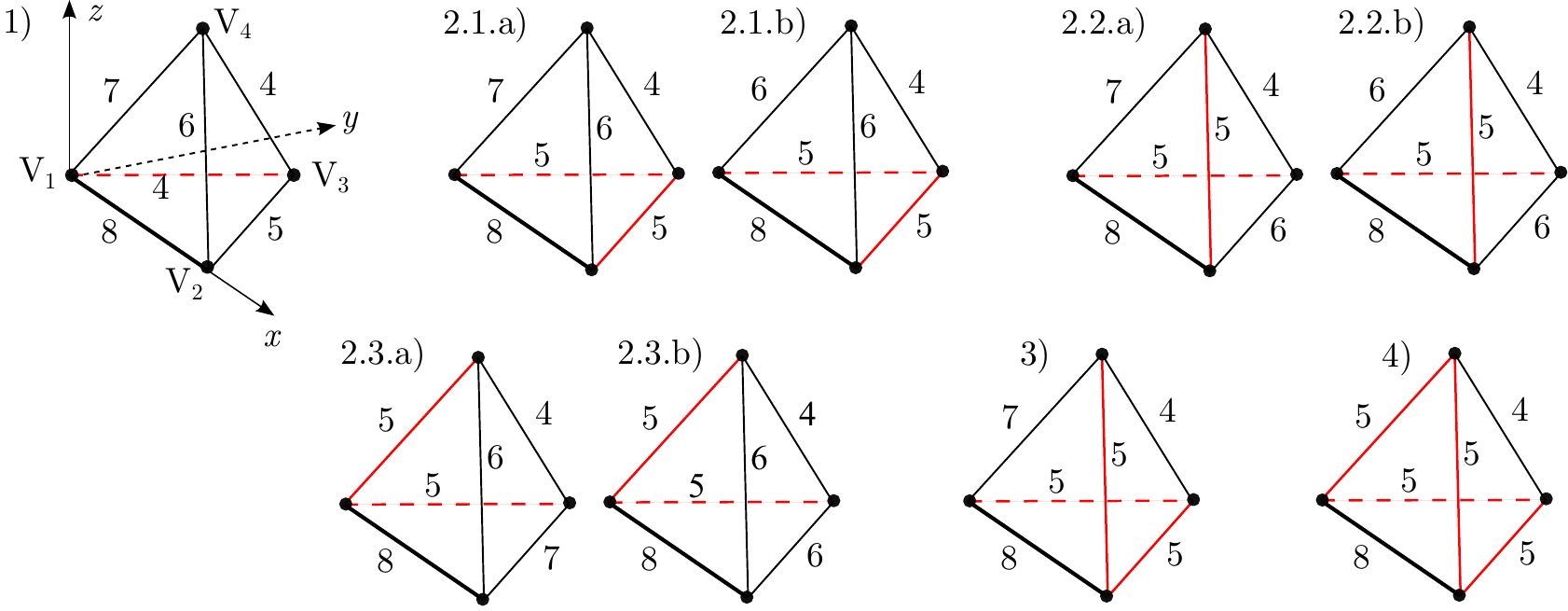}
  \caption{The nine edge configurations arising in the selection of the canonical face. The  edges of minimum length adjacent to the unique longest edge are highlighted in red, while the unique longest edge is drawn as a thick black line.}
  \label{fig:8_cases}
\end{figure}

The counterexample presented in Section \ref{sec:counterexample-uniqueness} shows that the normalization does not necessarily produce a unique normalized coordinate triple. This may occur when two or more edges adjacent to the selected longest edge have the same minimum length and remain admissible under the normalization criteria. We now analyze this phenomenon systematically.
In this section, we restrict our analysis to tetrahedra with a unique longest edge $V_1V_2$, as shown in Figure~\ref{fig:8_cases}. 

The four edges adjacent to the longest edge are $V_1V_3,V_1 V_4,V_2V_3$, and $V_2V_4$. 
The normalization procedure selects an edge of minimum length among these four in order to determine the canonical face and its labeling. Therefore, under the assumption of a unique longest edge, ambiguity may arise only when two or more of these adjacent edges have the same minimum length.
The possible configurations can then be classified according to the number and arrangement of the competing edges of minimum length.

Since the unique longest edge has exactly four adjacent edges, the minimum length can be attained by one, two, three, or all four. When the minimum length is attained by exactly two adjacent edges, there are, up to vertex relabeling, three distinct ways in which they can be arranged. These give rise to Cases 2.1, 2.2, and 2.3. Each of these three cases may further split into two subcases, depending on additional equality relations among the remaining edges. Together with the cases in which the minimum length is attained by one, three, or all four adjacent edges, this yields the nine configurations illustrated in Figure~\ref{fig:8_cases}.

These nine configurations provide all possible arrangements relevant to the determination of the canonical face and its labeling.

\subsection{Unique edge of minimum length. Case 1}

This configuration is illustrated in Figure~\ref{fig:8_cases}.1. Hence, no tie occurs and the canonical face $V_1V_2V_3$ is uniquely determined by the length criterion. 
A representative example is given by the following similarity class and its corresponding triple:
\[
[T_1]=[8,7,6,5,4,4]
\longleftrightarrow
\tau_1=
\left(
\frac{7}{16}+\frac{\sqrt{79}}{16}i,\left(
\frac{9}{16}+\frac{49\sqrt{79}}{1264}i,
\frac{\sqrt{10981}}{158}\right)
\right).
\]

No additional angle or symmetry criterion is required. Any tetrahedron with sextuple \((A,B,C,D,E,F)\) belongs to this case whenever the minimum length is attained uniquely by one of \(C\), \(D\), or \(E\). 

\subsection{Two  edges of the same minimum length. Case 2}

When two of the four edges adjacent to the longest edge  attain the same minimum length, three distinct combinatorial arrangements are possible.

\subsubsection{The two edges of minimum length lie on the same face sharing the longest edge. Case 2.1}

The canonical face is the isosceles triangle $V_1V_2V_3$. Two situations arise. 

\smallskip
\noindent\textbf{Asymmetric Case 2.1.a}

This case has already been studied in detail in Section \ref{sec:counterexample-uniqueness} and is illustrated in  Figures~\ref{fig:fallo-korotov} and~\ref{fig:8_cases}.1.a. Tetrahedra in this case have sextuples of the form
$(A,B,C,D,D,F)$ with $C>D$.

\smallskip
\noindent\textbf{Symmetric Case 2.1.b}

If the face, $V_1V_2V_4$, 
is also isosceles, as shown in Figure~\ref{fig:8_cases}.1.b, the tetrahedron has an additional reflection symmetry. 
Consequently, the two admissible labelings lead to the same normalized coordinate representation, although neither the length criterion nor the angle criterion distinguishes between them.

The following similarity class provides a representative example, with its associated triple:
\[
[T_3]=[8,6,6,5,5,4]\;\longleftrightarrow\;
\tau_3=
\left(
\frac12+\frac{\sqrt6}{4}i,\left(
\frac12+\frac{\sqrt6}{8}i,
\frac{\sqrt{26}}{8}\right)
\right).
\]
Tetrahedra in this case have sextuples of the form $(A,B,B,D,D,F)$.

\subsubsection{The two edges of minimum length are opposite. Case 2.2}

\smallskip
\noindent\textbf{Unequal remaining adjacent edges. Case 2.2.a}

The canonical face, $V_1V_2V_3$, is uniquely determined by the angle criterion, since $\angle V_2V_1V_3 < \angle V_1V_2V_4$, as shown in Figure~\ref{fig:8_cases}.2.a.  Hence, the normalization is unique.
The following similarity class and its associated triple provide this example:
\[
[T_4]=[8,7,5,6,5,4]\;\longleftrightarrow\;
\tau_4=
\left(
\frac{7}{16}+\frac{\sqrt{111}}{16}i,\left(
\frac58+\frac{29\sqrt{111}}{888}i,
\frac{5\sqrt{2886}}{444}\right)
\right).
\]
Tetrahedra in this case have sextuples of the form 
$(A,B,C,D,C,F)$, with $ C<D$.

\smallskip
\noindent\textbf{Equal remaining adjacent edges. Case 2.2.b}

 This case is shown in Figure~\ref{fig:8_cases}.2.b. The tetrahedron has a half-turn symmetry that interchanges the two admissible edges of minimum length. Accordingly, the canonical face can be either \(V_1V_2V_3\) or \(V_1V_2V_4\), and both possible choices lead to the same normalized coordinate representation after the corresponding vertex relabeling.
The following similarity class and its corresponding triple provide a representative example:
\[[T_5]=[8,6,5,6,5,4]\;\longleftrightarrow\;
\tau_5=
\left(
\frac{7}{16}+\frac{\sqrt{111}}{16}i,\left(
\frac9{16}+\frac{49\sqrt{111}}{1776}i,
\frac{\sqrt{17205}}{222}\right)
\right).
\]

Tetrahedra in this case have sextuples of the form $(A,B,C,B,C,F)$.

\subsubsection{Two edges of minimum length lie on a face sharing the edge opposite to the longest edge. Case 2.3}

\smallskip
\noindent\textbf{Unequal opposite edges. Case 2.3.a}

The angle criterion resolves the tie between the two admissible choices for the canonical face $V_1V_2V_3$ or $V_1V_2V_4$,
 since $\angle V_2V_1V_4 < \angle V_2V_1V_3$. Thus, the selected  canonical face is $V_1V_2V_4$, as illustrated in Figure~\ref{fig:8_cases}.3.a. Hence, the normalized coordinate representation is unique.
An example of this case is provided by the following similarity class together with its triple:
\[
[T_6]=[8,7,5,5,6,4]\;\longleftrightarrow\;
\tau_6=
\left(
\frac{7}{16}+\frac{\sqrt{111}}{16}i,\left(
\frac38+\frac{9\sqrt{111}}{148}i,
\frac{\sqrt{8362}}{148}\right)
\right).
\]

Tetrahedra in this case have sextuples of the form
$(A,B,C,C,E,F)$, with $C<E.$

\smallskip
\noindent\textbf{ Equal opposite edges. Case 2.3.b}

The
canonical face can be either $V_1V_2V_3$ or $V_1V_2V_4$, 
and both faces yield the same normalized coordinate representation after an appropriate relabeling of the vertices, as shown in Figure~\ref{fig:8_cases}.3.b.  
This configuration is illustrated by the following similarity class and its triple: 
\[
[T_7]=[8,6,5,5,6,4]\;
\longleftrightarrow\;
\tau_7=
\left(
\frac7{16}+\frac{\sqrt{111}}{16}i,\left(
\frac7{16}+\frac{47\sqrt{111}}{1776}i,
\frac{\sqrt{17538}}{222}\right)
\right).
\]

Unlike Cases 
2.1.b and 2.2.b, uniqueness here does not arise from a geometric symmetry of the tetrahedron, but from the fact that the two admissible choices are equivalent under vertex relabeling.
Tetrahedra in this case have sextuples of the form
$(A,B,C,C,B,F)$.

\subsection{Three edges of the same minimum length. Case 3}

The angle criterion distinguishes between the competing canonical faces, selecting $V_1V_2V_3$ rather than $V_1V_2V_4$. However, the resulting canonical face is again isosceles. As a result, the remaining ambiguity is of the same type as that occurring in Case 2.1.a, producing two distinct normalized coordinate representations, as shown in Figure~\ref{fig:8_cases}.3.

This case is illustrated by the following similarity class, which gives rise to two distinct triples: 
\[
[T_8]=[8,7,5,5,5,4]
\;\longleftarrow\;
\left\{
\begin{gathered}
\tau_8=
\left(
\frac12+\frac{\sqrt6}{4}i,\left(
\frac58+\frac{\sqrt6}{8}i,
\frac58\right)
\right)
\; \text{for the choice } V_1V_3,
\\[0.25mm]
\text{and}
\\[0.25mm]
\tau_8'=
\left(
\frac12+\frac{\sqrt6}{4}i,\left(
\frac38+\frac{\sqrt6}{8}i,
\frac58\right)
\right)
\; \text{for the choice } V_2V_3.
\end{gathered}
\right.
\]
Once again, $\mathcal R_{\tau}(\tau_8)=\tau'_8$ and therefore the two distinct triples 
$\tau_8$ and $\tau_8'$  correspond to the same similarity class.
Thus, Case 
3 also produces  non-uniqueness. Tetrahedra  in this case have sextuples of the form 
$(A,B,C,C,C,F)$.

\subsection{Four edges of the same minimum length. Case 4}

This configuration is highly symmetric, as shown in  Figure~\ref{fig:8_cases}.4.  
The canonical face can be either \(V_1V_2V_3\) or \(V_1V_2V_4\), and each canonical face admits two admissible labelings. All of them lead to the same normalized coordinate representation.

The following similarity class and its corresponding triple provide this example:
\[
[T_9]=[8,5,5,5,5,4]\;
\longleftrightarrow\;
\tau_9=
\left(
\frac1{2}+\frac{\sqrt{6}}{4}i,\left(
\frac1{2}+\frac{\sqrt{6}}{12}i,
\frac{\sqrt{3}}{3}\right)
\right).
\]

The tetrahedron $T_9$ admits two reflection symmetries. Their composition is a rotation, so the different admissible choices are geometrically equivalent.
Hence, the normalized coordinate representation is unique.
Tetrahedra in this case have sextuples of the form
$(A,B,B,B,B,F)$.

Note that, in all nine cases, the edge \(F\), opposite  the longest edge \(A\), is irrelevant to the selection of the edge of minimum length among those adjacent to the longest edge and, consequently, to the determination of the canonical face. However, the condition \(F<A\) must be satisfied.

The previous analysis can now be summarized in a single Theorem.

\begin{theorem}
 For tetrahedra with a unique longest edge, among the nine possible edge configurations arising in the determination of the canonical face, non-uniqueness of the Michaud–Korotov normalization occurs in Cases 2.1.a and 3.
In the remaining seven configurations, the normalized representative is uniquely determined for one of the following reasons:
\[
\begin{array}{c|c}
\text{Case} & \text{Mechanism ensuring uniqueness}\\
\hline
1 & \text{unique adjacent edge of minimum length}\\
2.1.b & \text{reflection symmetry}\\
2.2.a & \text{angle  criterion}\\
2.2.b & \text{half-turn symmetry}\\
2.3.a & \text{angle criterion}\\
2.3.b & \text{equivalent vertex relabeling}\\
4 & \text{geometric symmetry}
\end{array}
\]

Consequently, equality among the edges of minimum length adjacent to the longest edge does not by itself cause non-uniqueness. Non-uniqueness occurs only when two or more admissible choices remain under the normalization criteria and these choices lead to different normalized coordinate representations.
\end{theorem}


It should be noted that non-uniqueness may also arise when the tetrahedron has more than one longest edge, since different admissible choices of the longest edge may lead to different normalized coordinate representations. The analysis of this case is straightforward and therefore omitted for brevity.

\section{Overcounting in the LEB orbits}

The lack of similarity invariance is not merely a representational issue. It directly affects the cardinality of normalized LEB orbits and, consequently, the characterization of LEB convergence.

As shown in Sections~\ref{sec:counterexample-uniqueness} and~\ref{sec:classification}, if two reflected  triples, $\tau$ and $\tau'$, are stored as distinct points of the coordinate space, they are counted as different orbit elements even though they correspond to the same similarity class. Thus, the orbit in normalized-coordinate space contains more points than the corresponding orbit in similarity-class space.

This leads to the following proposition concerning the perturbed Sommerville tetrahedron considered in \cite{michaud2026orbits}.

\begin{proposition}[Perturbed Sommerville orbit count]\label{prop:sommerville}
The LEB orbit of the perturbed Sommerville class $[401,321,300,300,321,400]$ contains 21 distinct similarity classes  whereas the normalized-coordinate orbit reported in~\cite{michaud2026orbits} contains 25 triples.
\end{proposition}

\begin{proof}

Consider the perturbed Sommerville tetrahedron introduced in Example 3.1 of \cite{michaud2026orbits}, with vertices
$(-1,0,0),\; (1,0,0),\; (0,-1,1),$ and $\left(0,1,\tfrac{11}{10}\right)$, which belongs to the similarity class $[401,321,300,300,321,400].$

Michaud and Korotov report an orbit containing \(25\) triples. However, several pairs within the reported orbit correspond to the same similarity class, thereby leading to an overcount of the orbit cardinality.

Note that \(\tau_3\) and \(\tau_5\) appear separately in the orbit reported in Example 3.1 of \cite{michaud2026orbits}:
\[
\tau_3=
\left(\frac{1}{2}+\frac{10}{21}i,
\left(\frac{10}{21}+\frac{10}{21}i,\frac{10}{21}\right)\right)\;\text{ and }\;
\tau_5=
\left(\frac{1}{2}+\frac{10}{21}i,
\left(\frac{11}{21}+\frac{10}{21}i,\frac{10}{21}\right)\right).
\]


These two triples are related by reflection across the plane $x=1/2$, since 
$\mathcal R_{\tau}(\tau_3)=\tau_5$. 
It is straightforward to verify that only the $x$-coordinate of the fourth vertex differs, which is $\frac{10}{21}$ for $\tau_3$ and $\frac{11}{21}$ for $\tau_5$. Therefore, the two reflected triples are distinct but correspond to the same similarity class $[1764,1284,1200,841,841,401]$.
 

Similarly, three additional pairs among the 25 triples reported in~\cite{michaud2026orbits} consist of reflected triples corresponding to the same tetrahedral similarity class:
\[
\begin{aligned}
&\mathcal R_{\tau}(\tau_6)=\tau_{10},\quad\;\, \{\tau_6,\tau_{10}\}\,\,
\longrightarrow
[1604,1284,1200,801,801,441],\\
&\mathcal R_{\tau}(\tau_{21})=\tau_{22},\quad \{\tau_{21},\tau_{22}\}
\longrightarrow
[800,441,401,300,300,321],\\
&\mathcal R_{\tau}(\tau_{24})=\tau_{25},\quad \{\tau_{24},\tau_{25}\}
\longrightarrow
[884,441,401,321,321,300].
\end{aligned}
\]
Together with the pair  $\{\tau_3,\tau_5\}$, these four reflected pairs account for four duplicate entries in the normalized-coordinate orbit. Hence, the 25 reported triples represent
$25-4=21$
distinct similarity classes. The genealogy tree obtained with SCLEB confirms 
this count, as shown in Appendix A.
\end{proof}

\begin{remark} Note that the perturbed Sommerville tetrahedron in the similarity class
$[401,321,300,300,321,400]$
has a generalized sextuple of the form (A,B,C,C,B,F) and belongs to the family \(R_1^+\). According to \cite{PadronTrujilloSuarez2025}, after identifying configurations belonging to the same similarity class, its orbit contains 21 distinct similarity classes.  
\end{remark}

\begin{lemma}
The normalized LEB orbit  obtained using the  Michaud--Korotov normalization \cite{michaud2026orbits}  overcounts tetrahedral  similarity classes in some cases.
\end{lemma}

\begin{proof}
Let $T$ and $T'$ be two reflected tetrahedra. Since a reflection is an orthogonal transformation, $T\sim T'$. However, if the normalization assigns distinct coordinate triples,
$
\mathcal N(T)\neq \mathcal N(T'),
$
then the same similarity class is represented by two distinct elements of the normalized LEB orbit. Hence,
$
\left|\mathcal O_{\mathcal N}(T_0)\right|>
\left|\mathcal O_{\mathrm{LEB}}([T_0])\right|
$
may occur.
\end{proof}





As a consequence of the overcounting that may occur when LEB orbits are represented using the normalization in \cite{michaud2026orbits}, the following remarks hold.

\begin{remark}

Conjecture~3.1 of~ \cite{michaud2026orbits}  reports orbit lengths in $\{7,8,21,22,23,24,25,26\}$. However, SCLEB yields orbit length $20$ for   tetrahedra in the families \(\widetilde{T}^{H}_{1,\alpha}\) through \(\widetilde{T}^{H}_{4,\alpha}\), and orbit length $7$  for the tetrahedra in the family \(\widetilde{T}^{H}_{5,\alpha}\). 
This discrepancy further illustrates the sensitivity of the computed orbit cardinality to the normalization procedure.

\end{remark}

\begin{remark}
The orbit lengths $\{39,37,43,43,43,37\}$ reported in Theorem~3.2
of \cite{michaud2026orbits}  do not represent the corresponding numbers of distinct
similarity classes for the tetrahedra $T_1,\ldots,T_6$. These tetrahedra, introduced in
\cite{SuTrTa}, belong to the family $R_1^+$, as proved in \cite{PadronTrujilloSuarez2025,TrSuPa}. Using the squared-edge sextuple representation, all six tetrahedra
have LEB orbits  containing $37$ distinct similarity classes.
Thus, the values $39$ and $43$ reported in \cite{michaud2026orbits}
are inconsistent with the maximum value of $37$ for the $R_1^+$ family. Table~\ref{table:37} identifies the reflected triples  corresponding to the same tetrahedral similarity class for the tetrahedron \(T_1\).
\end{remark}

\section{Final remarks}

This work has examined the non-uniqueness of a coordinate-based normalization of tetrahedral similarity classes and its consequences for the analysis of Longest-Edge Bisection (LEB) orbits. 
In particular, we have shown that this normalization is not similarity-invariant in general: two similar tetrahedra may be assigned distinct normalized coordinate representations. Consequently, the normalization does not induce a well-defined map on similarity classes in general, and normalized orbit states cannot necessarily be identified with distinct tetrahedral similarity classes.

For tetrahedra with a unique longest edge, the possible ambiguities in the determination of the canonical face and its labeling have been classified into nine edge configurations. Among these configurations, non-uniqueness occurs in Cases~2.1.a and~3, whereas in the remaining cases the admissible choices lead to a unique normalized coordinate representation, either through the selection criteria or through geometric symmetry and vertex relabeling.

This distinction has direct consequences for orbit counting. In particular, for the perturbed Sommerville tetrahedron considered in \cite{michaud2026orbits}, the normalized-coordinate orbit contains 25 triples, whereas these triples represent only 21 distinct similarity classes. More generally, whenever different normalized representations correspond to the same similarity class, the cardinality of the normalized LEB orbit may overestimate the cardinality of the corresponding orbit in similarity-class space.

A separate and equally important issue arises when a tetrahedron has more than one longest edge. In this situation,
the ambiguity no longer concerns only the representation of a similarity class: the choice of the edge to be bisected is part of the subdivision rule itself rather than merely a representational issue. Different admissible choices may generate different descendants and therefore different LEB orbits. This distinction is particularly relevant  for the regular tetrahedron, where the adopted tie-breaking rule may substantially affect the subsequent subdivision process.

Thus, two conceptually different sources of ambiguity must be distinguished in the study of LEB: the representation of tetrahedra belonging to the same similarity classes, investigated in the present work, and the selection of the edge to be bisected when several longest edges are available. 
The latter problem will  be investigated separately, together with the construction of a canonical coordinate representation compatible with a deterministic longest-edge selection rule.

The results presented here provide  a more precise mathematical framework for the study of tetrahedral LEB dynamics, in which the subdivision process,  the identification of similarity classes, and their coordinate representation are treated as distinct but closely related components.

\AddToShipoutPictureFG{%
  \ifnum\value{page}=\getpagerefnumber{page:genealogy-tree}
    \AtPageLowerLeft{%
       \put(\LenToUnit{0.52\paperwidth},\LenToUnit{0.23\paperheight}){%
       \makebox(0,0){%
        \color{white}\rule{0.6cm}{0.6cm}%
       }%
      }%

      \put(\LenToUnit{0.800\paperwidth},\LenToUnit{0.53\paperheight}){%
        \makebox(0,0){\rotatebox{90}{\small\thepage}}%
      }%

      \put(\LenToUnit{0.779\paperwidth},\LenToUnit{0.555\paperheight}){%
        \makebox(0,0){%
          \rotatebox{90}{%
            \small
            Triples $\tau_2, \tau_3, \tau_4, \tau_5, \tau_6, \tau_9, \tau_{10}, \tau_{11},\tau_{21}, \tau_{22}$ and $\tau_{24}$ 
            belong to Case 2.1, whereas $\tau_1$ corresponds to Case 2.3.b; see  Figure~\ref{fig:8_cases}.
          }%
        }%
      }%
    }%
  \fi
}

\appendix
\refstepcounter{section}
\addcontentsline{toc}{section}{Appendix \thesection. Genealogy tree for a LEB class \texorpdfstring{$\bm{\tau_1\leftrightarrow[401,321,300,300,321,400]}$}
{tau1 - [401,321,300,300,321,400]}}

\begin{center}
\begin{tikzpicture}[rotate=90,transform shape,
scale=0.820,
  >=Stealth,
  every node/.style={font=\fontsize{10.75}{12.25}\selectfont},
  edge/.style={->, thick}
]
\label{page:genealogy-tree}
\node[overlay,rotate=90,transform shape=false,font=\normalfont\small\bfseries,align=center] at (-1.0,1.5) {Appendix \thesection. Genealogy tree for a LEB class $\bm{\tau_1\leftrightarrow[401,321,300,300,321,400]$}};
\node at (-12.75,-1.5) {Iter 1};
\node at (-12.75,-3) {Iter 2};
\node at (-12.75,-4.5) {Iter 3};
\node at (-12.75,-6) {Iter 4};
\node at (-12.75,-7.5) {Iter 5};
\node at (-12.75,-9) {Iter 6};
\node at (-12.75,-10.5) {Iter 7};

  \node (T1)  at (0, 0){$\tau_{1}\leftrightarrow[401,321,300,300,321,400]$};
  \node[red] (T2)  at (-4.6,-1.5){$\tau_{11}\leftrightarrow[1600,1284,1284,841,841,401]$};
  \node[blue] (T3)  at ( 5.0,-1.5){$\tau_{2}\leftrightarrow[1600,1200,1200,841,841,401]$};
  \node[violet] (T4)  at (-4.6,-3){$\tau_{15}\leftrightarrow[1284,884,400,841,401,841]$};
  \node[orange] (T6)  at ( 5.0,-3){$\tau_{13}\leftrightarrow[1200,841,401,800,400,441]$};
  \node[cyan] (T8)  at (-9.1,-4.5){$\tau_{18}\leftrightarrow[841,441,400,321,300,321]$};
  \node[green] (T9)  at (-3.45,-4.5){$\tau_{24}\leftrightarrow[884,441,401,321,321,300]$};
  \node[green, rotate=-90] (eqTau24) at ($(T9.south)+(-2.50,-0.10)$) {$\ne$};
  \node[green] at ($(eqTau24.south)+(0.16,-0.30)$) {$\tau_{25}$};
  \node[brown] (T13) at (2.25,-4.5){$\tau_{21}\leftrightarrow[800,441,401,300,300,321]$};
  \node[brown, rotate=90] (eqTau21) at ($(T13.south)+(-2.50,-0.1)$) {$\ne$};
  \node[brown] at ($(eqTau21.south)+(-0.29,-0.3)$) {$\tau_{22}$};
  \node[magenta] (T12) at (7.9,-4.5){$\tau_{20}\leftrightarrow[841,441,400,300,321,300]$};
  \node[teal] (T16) at (-7.1,-6){$\tau_{3}\leftrightarrow[1764,1284,1200,841,841,401]\leftrightarrow\tau_{5}$};
  \node[NavyBlue] (T17) at (-0.5,-6){$\tau_{7}\leftrightarrow[401,321,300,200,221,100]$};
  \node[ForestGreen] (T20) at ( 5.75,-6){$\tau_{8}\leftrightarrow[1764,1284,1200,800,884,400]$};
  \node[Sepia] (T22) at (-9.2,-7.5) {$\tau_{17}\leftrightarrow[1284,841,401,801,441,400]$};
  \node[RawSienna] (T23) at (-3.3,-7.5){$\tau_{14}\leftrightarrow[1200,841,401,801,441,400]$};
  \node[BrickRed] (T24) at ( 2.55,-7.5) {$\tau_{16}\leftrightarrow[1284,884,400,801,441,401]$};
  \node[draw=red, fill=red, text=white, rounded corners=2pt, inner sep=2pt] (T30) at ( 8.5,-7.5) {$\tau_{12}\leftrightarrow[1200,801,441,800,400,401]$};
  \node[Periwinkle] (T33) at (-7.0,-9){$\tau_{23}\leftrightarrow[801,401,400,300,321,300]$};
  \node[CadetBlue] (T31) at (-0.10,-9) {$\tau_{19}\leftrightarrow[801,401,400,321,300,321]$};
  \node[draw=blue, fill=blue, text=yellow, rounded corners=2pt, inner sep=2pt] (T34) at (-2.35,-10.5) {$\tau_{6}\leftrightarrow[1604,1284,1200,801,801,441]\leftrightarrow\tau_{10}$};
  \node[Salmon] (T35) at ( -9.05,-10.5) {$\tau_{9}\leftrightarrow[1600,1200,1200,801,801,441]$};
  \node[draw=MidnightBlue, fill=MidnightBlue, text=white, rounded corners=2pt, inner sep=2pt] (T37) at (4.5,-10.5) {$\tau_{4}\leftrightarrow[1600,1284,1284,801,801,441]$};

  \coordinate (T1split) at ($(T1.south)+(0,0.05)$);
  \draw[edge] (T1split) -- (T2);
  \draw[edge] (T1split) -- (T3);

  \draw[edge] (T2) -- (T4);
  \draw[edge] (T3) -- (T6);
  \coordinate (T4split) at ($(T4.south)+(0,0.05)$);
  \draw[edge] (T4split) -- (T8);
  \draw[edge] (T4split) -- ([xshift=1.20cm]T9.north);

  \coordinate (T6split) at ($(T6.south)+(0,0.05)$);
  \coordinate (T6split2) at ([xshift=0.5cm]T6split);
  \draw[edge] (T6split2) -- (T12);
  \draw[edge] (T6split2) -- (T13);
%
%
  \draw[edge] ([xshift=0.15cm]T8.south) -- ([xshift=-1.85cm]T16.north);

  \coordinate (T13split) at ($(T13.south)+(0,0.05)$);
  \coordinate (T13split2) at ([xshift=0.35cm]T13split);
  \draw[edge] (T13split2) -- (T17);
  \draw[edge] (T13split2) -- (T20);
  \draw[edge] ([xshift=1.55cm]T9.south) -- ([xshift=-1.4cm]T17.north);
%
%
  \draw[edge] ([xshift=0.18cm]T9.south) -- ([xshift=0.18cm,yshift=-1.75cm]T9.south) -| ([xshift=0.75cm,yshift=-0.13cm]T20.west);

  \draw[edge] ([xshift=-0.45cm]T8.north) -- ([xshift=-0.45cm,yshift=2.67cm]T8.north) -- ([xshift=-6.3cm]T2.east);
  
  \draw[edge] ([xshift=0.65cm]T12.north) -- ([xshift=0.65cm,yshift=2.68cm]T12.north) -- ([xshift=6.15cm]T3.west);

  \coordinate (T16split) at ($(T16.south)+(0,0.05)$);
  \coordinate (T16split2) at ([xshift=0.7cm]T16split);
  \draw[edge] (T16split2) -- (T22);
  \draw[edge] (T16split2) -- (T23);
  \coordinate (T17up) at ([xshift=-0.1cm,yshift=2.68cm]T17.north);
  \draw[edge,-] ([xshift=-0.1cm]T17.north) -- (T17up);
  \draw[edge] (T17up) -- ([xshift=-0.10cm]T4.east);
  \draw[edge] (T17up) -- ([xshift=0.10cm, yshift=-0.04cm]T6.west);
  \draw[edge] ([xshift=-0.25cm]T23.north) -- ([xshift=-0.25cm,yshift=1.62cm]T23.north) -| ([xshift=-0.75cm,yshift=-0.25cm]T8.east);
  \coordinate (T20split) at ($(T20.south)+(0,0.05)$);
  \coordinate (T20split2) at ([xshift=0.1cm]T20split);
  \draw[edge] (T20split2) -- (T24);
  \draw[edge] (T20split2) -- (T30);
%
%
  \draw[edge] ([xshift=-1.55cm]T24.north) -- ([xshift=-1.55cm,yshift=0.4cm]T24.north) -- ([xshift=0.02cm,yshift=-1.93cm]T9.south) -- ([xshift=0.02cm]T9.south);
%
%
  \draw[edge] ([xshift=-1.55cm]T24.south) -- ([xshift=1.10cm]T31.north);
  \draw[edge] ([xshift=0.60cm]T30.north) -- ([xshift=0.60cm,yshift=2.0cm]T30.north) -| ([xshift=-0.75cm,yshift=-0.20cm]T13.east);

  \draw[edge] ([xshift=0.9cm]T22.south) --([xshift=-1.30cm]T33.north); 

  \coordinate (T33split) at ($(T33.south)+(0,0.05)$);
  \draw[edge] (T33split) -- ([xshift=-2.5cm]T34.north);
  \draw[edge] (T33split) -- ([xshift=0.25cm]T35.north);
  \draw[edge] ([xshift=2.25cm]T23.south) -- ([xshift=-0.95cm]T31.north);
%
%
  \coordinate (T31split) at ($(T31.south)+(0.20,-0.05)$);
  \draw[edge] (T31split) -- ([xshift=-0.15cm]T34.north);
  \draw[edge] (T31split) -- ([xshift=-2.08cm]T37.north);
 %
 %
  \draw[edge] ([xshift=-1.25cm]T34.north) -- ([xshift=-1.25cm,yshift=1.70cm]T34.north) -- ([xshift=1.5cm,yshift=-0.7cm]T22.south) -- ([xshift=1.5cm]T22.south);
  \draw[edge] ([xshift=-0.88cm]T34.north)--([xshift=0.05cm]T23.south);
%
%
  \draw[edge] ([xshift=-2.15cm,yshift=-0.20cm]T22.north)-- ([xshift=-2.93cm,yshift=0.7cm]T22.north)-- ([xshift=-12.13cm,yshift=0.5cm]T1.north) -- ([xshift=1.7cm,yshift=4.85cm]T12.north)-- ([xshift=1.7cm]T12.north);
 
%
  \draw[edge] ([xshift=1.10cm]T12.north) -- ([xshift=1.10cm,yshift=4.55cm]T12.north) -- ([xshift=-11.99cm,yshift=0.20cm]T1.north) -- ([xshift=-4.91cm,yshift=-0.32cm]T16.north) -- ([xshift=-7.10cm]T16.east);
%
%
 \draw[edge] ([xshift=0.5cm]T30.south) -- ([xshift=0.5cm,yshift=-3.9cm]T30.south) -- ([xshift=-8.90cm,yshift=-3.9cm]T30.south) -- ([xshift=-3.10cm,yshift=-0.85cm]T35.south) -- ([xshift=-3.10cm,yshift=1.25cm]T35.south) -- ([xshift=0.10cm]T33.west);
 %
%
  \draw[edge] ([xshift=-0.96cm]T37.north)--([xshift=0.98cm]T24.south);
%
   \draw[edge] (T35.south) -- ++(0,-0.5) -- ++(17.53,0) -- ([xshift=-0.04cm]T30.south); 
\end{tikzpicture}
\end{center}
\thispagestyle{empty}
\begin{table}[h!]

\centering
\scriptsize

\setlength{\tabcolsep}{3pt}

\renewcommand{\arraystretch}{1.08}

\begin{tabular}{c@{}l@{\hspace{3pt}}c@{\hspace{3pt}}l}

\toprule

\multicolumn{2}{c}{Normalized quintuple} & & Similarity class \\

\midrule

$\tau_1\!\!=$ &
$\left(\frac{21}{76}+\frac{\sqrt{1079}}{76}i,\left(\frac{13}{19}+\frac{233\sqrt{1079}}{41002}i,\frac{\sqrt{260424203}}{41002}\right)\right)$
&
$\leftrightarrow$
&
$[76{,}54{,}20{,}50{,}22{,}29]\teq [T_{\scalebox{0.65}{$4$}}$]
\\[3.5pt]

$\tau_2\!\!=$ &
$\left(\frac{53}{108}+\frac{\sqrt{1079}}{108}i,\left(\frac{61}{108}+\frac{7\sqrt{1079}}{116532}i,\frac{\sqrt{82239222}}{19422}\right)\right)$
&
$\leftrightarrow$
&
$[54{,}29{,}22{,}19{,}18{,}17]\teq [T_{\scalebox{0.65}{$9$}}$]
\\[3.5pt]

$\tau_3\!\!=$ &
$\left(\frac{43}{88}+\frac{\sqrt{1759}}{88}i,\left(\frac{5}{11}+\frac{225\sqrt{1759}}{19349}i,\frac{\sqrt{491580694}}{38698}\right)\right)$
&
$\leftrightarrow$
&
$[88{,}76{,}68{,}41{,}43{,}29]\teq [T_{\scalebox{0.65}{$36$}}$]
\\[3.5pt]

$\tau_4\!\!=$ &
$\left(\frac{20}{41}+\frac{3\sqrt{33}}{41}i,\left(\frac{39}{82}+\frac{20\sqrt{33}}{4059}i,\frac{\sqrt{17187159}}{8118}\right)\right)$
&
$\leftrightarrow$
&
$[41{,}22{,}20{,}17{,}18{,}17]\teq [T_{\scalebox{0.65}{$30$}}$]
\\[3.5pt]

$\tau_5\!\!=$ &
$\left(\frac{55}{116}+\frac{\sqrt{2543}}{116}i,\left(\frac{27}{58}+\frac{1299\sqrt{2543}}{147494}i,\frac{\sqrt{936808141}}{73747}\right)\right)$
&
$\leftrightarrow$
&
$[58{,}38{,}34{,}24{,}27{,}10]\teq [T_{\scalebox{0.65}{$20$}}$]
\\[3.5pt]

$\tau_6\!\!=$ &
$\left(\frac{39}{80}+\frac{\sqrt{1759}}{80}i,\left(\frac{1}{2}+\frac{41\sqrt{1759}}{3518}i,\frac{\sqrt{111722885}}{17590}\right)\right)$
&
$\leftrightarrow$
&
$[80{,}68{,}68{,}43{,}41{,}29]\teq [T_{\scalebox{0.65}{$35$}}$]
\\[3.5pt]

$\tau_7\!\!=$ &
$\left(\frac{5}{17}+\frac{2\sqrt{15}}{17}i,\left(\frac{12}{17}+\frac{217\sqrt{15}}{4080}i,\frac{\sqrt{3239265}}{4080}\right)\right)$
&
$\leftrightarrow$
&
$[68{,}50{,}22{,}48{,}20{,}29]\teq [T_{\scalebox{0.65}{$6$}}$]
\\[3.5pt]

$\tau_8\!\!=$ &
$\left(\frac{12}{25}+\frac{\sqrt{274}}{50}i,\left(\frac{59}{100}-\frac{2\sqrt{274}}{3425}i,\frac{\sqrt{1740311}}{2740}\right)\right)$
&
$\leftrightarrow$
&
$[50{,}29{,}20{,}19{,}17{,}18]\teq [T_{\scalebox{0.65}{$8$}}$]
\\[3.5pt]

\textcolor{blue}{$\tau_9\!\!=$} &
\textcolor{blue}{$\left(\frac{1}{2}+\frac{\sqrt{15}}{12}i,\left(\frac{41}{96}+\frac{\sqrt{15}}{240}i,\frac{\sqrt{63515}}{480}\right)\right)$}
&
\textcolor{blue}{$\rightarrow$}
&
\textcolor{blue}{$[48{,}29{,}22{,}17{,}17{,}18]\teq [T_{\scalebox{0.65}{$14$}}$]}
\\[3.5pt]

$\tau_{10}\!\!=$ &
$\left(\frac{5}{17}+\frac{2\sqrt{15}}{17}i,\left(\frac{10}{17}+\frac{263\sqrt{15}}{4080}i,\frac{\sqrt{3239265}}{4080}\right)\right)$
&
$\leftrightarrow$
&
$[68{,}48{,}20{,}41{,}29{,}22]\teq [T_{\scalebox{0.65}{$25$}}$]
\\[3.5pt]

$\tau_{11}\!\!=$ &
$\left(\frac{39}{80}+\frac{\sqrt{1759}}{80}i,\left(\frac{19}{40}+\frac{939\sqrt{1759}}{70360}i,\frac{\sqrt{111722885}}{17590}\right)\right)$
&
$\leftrightarrow$
&
$[80{,}76{,}72{,}41{,}43{,}29]\teq [T_{\scalebox{0.65}{$37$}}$]
\\[3.5pt]

$\tau_{12}\!\!=$ &
$\left(\frac{47}{96}+\frac{\sqrt{1055}}{96}i,\left(\frac{13}{32}+\frac{29\sqrt{1055}}{33760}i,\frac{\sqrt{40204995}}{12660}\right)\right)$
&
$\leftrightarrow$
&
$[48{,}29{,}20{,}17{,}18{,}17]\teq [T_{\scalebox{0.65}{$15$}}$]
\\[3.5pt]

$\tau_{13}\!\!=$ &
$\left(\frac{39}{80}+\frac{\sqrt{2319}}{80}i,\left(\frac{1}{2}+\frac{55\sqrt{2319}}{4638}i,\frac{\sqrt{147291285}}{23190}\right)\right)$
&
$\leftrightarrow$
&
$[40{,}34{,}34{,}25{,}24{,}11]\teq [T_{\scalebox{0.65}{$3$}}$]
\\[3.5pt]

$\tau_{14}\!\!=$ &
$\left(\frac{55}{116}+\frac{\sqrt{2543}}{116}i,\left(\frac{14}{29}+\frac{622\sqrt{2543}}{73747}i,\frac{\sqrt{936808141}}{73747}\right)\right)$
&
$\leftrightarrow$
&
$[58{,}36{,}34{,}24{,}27{,}10]\teq [T_{\scalebox{0.65}{$18$}}$]
\\[3.5pt]

$\tau_{15}\!\!=$ &
$\left(\frac{57}{116}+\frac{\sqrt{2319}}{116}i,\left(\frac{27}{58}+\frac{1187\sqrt{2319}}{134502}i,\frac{\sqrt{854289453}}{67251}\right)\right)$
&
$\leftrightarrow$
&
$[58{,}38{,}34{,}24{,}25{,}11]\teq [T_{\scalebox{0.65}{$21$}}$]
\\[3.5pt]

\textcolor{blue}{$\tau_{16}\!\!=$} &
\textcolor{blue}{$\left(\frac{1}{2}+\frac{\sqrt{15}}{12}i,\left(\frac{55}{96}+\frac{\sqrt{15}}{240}i,\frac{\sqrt{63515}}{480}\right)\right)$}
&
\textcolor{blue}{$\rightarrow$}
&
\textcolor{blue}{$[48{,}29{,}22{,}17{,}17{,}18]\teq [T_{\scalebox{0.65}{$14$}}$]}
\\[3.5pt]

$\tau_{17}\!\!=$ &
$\left(\frac{21}{44}+\frac{\sqrt{1055}}{44}i,\left(\frac{5}{11}+\frac{49\sqrt{1055}}{11605}i,\frac{\sqrt{294836630}}{23210}\right)\right)$
&
$\leftrightarrow$
&
$[22{,}19{,}17{,}17{,}18{,}20]\teq [T_{\scalebox{0.65}{$1$}}$]
\\[3.5pt]

\textcolor{violet}{$\tau_{18}\!\!=$} &
\textcolor{violet}{$\left(\frac{1}{2}+\frac{\sqrt{15}}{12}i,\left(\frac{41}{96}-\frac{\sqrt{15}}{240}i,\frac{\sqrt{63515}}{480}\right)\right)$}
&
\textcolor{violet}{$\rightarrow$}
&
\textcolor{violet}{$[48{,}29{,}22{,}17{,}17{,}19]\teq [T_{\scalebox{0.65}{$13$}}$]}
\\[3.5pt]

$\tau_{19}\!\!=$ &
$\left(\frac{23}{72}+\frac{\sqrt{1055}}{72}i,\left(\frac{7}{12}+\frac{97\sqrt{1055}}{12660}i,\frac{\sqrt{26803330}}{12660}\right)\right)$
&
$\leftrightarrow$
&
$[72{,}48{,}22{,}41{,}29{,}20]\teq [T_{\scalebox{0.65}{$22$}}$]
\\[3.5pt]

$\tau_{20}\!\!=$ &
$\left(\frac{41}{86}+\frac{\sqrt{1243}}{86}i,\left(\frac{45}{86}-\frac{39\sqrt{1243}}{106898}i,\frac{\sqrt{678962647}}{53449}\right)\right)$
&
$\leftrightarrow$
&
$[43{,}22{,}20{,}19{,}17{,}18]\teq [T_{\scalebox{0.65}{$32$}}$]
\\[3.5pt]

$\tau_{21}\!\!=$ &
$\left(\frac{5}{17}+\frac{2\sqrt{15}}{17}i,\left(\frac{41}{68}+\frac{287\sqrt{15}}{4080}i,\frac{\sqrt{3239265}}{4080}\right)\right)$
&
$\leftrightarrow$
&
$[68{,}48{,}20{,}43{,}29{,}22]\teq [T_{\scalebox{0.65}{$27$}}$]
\\[3.5pt]

$\tau_{22}\!\!=$ &
$\left(\frac{6}{19}+\frac{\sqrt{274}}{38}i,\left(\frac{45}{76}+\frac{315\sqrt{274}}{20824}i,\frac{\sqrt{66131818}}{20824}\right)\right)$
&
$\leftrightarrow$
&
$[76{,}50{,}22{,}43{,}29{,}20]\teq [T_{\scalebox{0.65}{$29$}}$]
\\[3.5pt]

$\tau_{23}\!\!=$ &
$\left(\frac{47}{96}+\frac{\sqrt{1055}}{96}i,\left(\frac{19}{32}-\frac{29\sqrt{1055}}{33760}i,\frac{\sqrt{40204995}}{12660}\right)\right)$
&
$\leftrightarrow$
&
$[48{,}29{,}20{,}18{,}17{,}19]\teq [T_{\scalebox{0.65}{$11$}}$]
\\[3.5pt]

$\tau_{24}\!\!=$ &
$\left(\frac{5}{17}+\frac{\sqrt{274}}{34}i,\left(\frac{10}{17}+\frac{331\sqrt{274}}{18632}i,\frac{\sqrt{59170574}}{18632}\right)\right)$
&
$\leftrightarrow$
&
$[68{,}50{,}22{,}41{,}29{,}20]\teq [T_{\scalebox{0.65}{$23$}}$]
\\[3.5pt]

$\tau_{25}\!\!=$ &
$\left(\frac{39}{80}+\frac{\sqrt{2319}}{80}i,\left(\frac{19}{40}+\frac{1219\sqrt{2319}}{92760}i,\frac{\sqrt{147291285}}{23190}\right)\right)$
&
$\leftrightarrow$
&
$[40{,}38{,}36{,}24{,}25{,}11]\teq [T_{\scalebox{0.65}{$2$}}$]
\\[3.5pt]

$\tau_{26}\!\!=$ &
$\left(\frac{41}{86}+\frac{\sqrt{1243}}{86}i,\left(\frac{41}{86}+\frac{39\sqrt{1243}}{106898}i,\frac{\sqrt{678962647}}{53449}\right)\right)$
&
$\leftrightarrow$
&
$[43{,}22{,}20{,}17{,}19{,}17]\teq [T_{\scalebox{0.65}{$33$}}$]
\\[3.5pt]

$\tau_{27}\!\!=$ &
$\left(\frac{12}{25}+\frac{\sqrt{274}}{50}i,\left(\frac{41}{100}+\frac{2\sqrt{274}}{3425}i,\frac{\sqrt{1740311}}{2740}\right)\right)$
&
$\leftrightarrow$
&
$[50{,}29{,}20{,}17{,}19{,}17]\teq [T_{\scalebox{0.65}{$12$}}$]
\\[3.5pt]

$\tau_{28}\!\!=$ &
$\left(\frac{53}{108}+\frac{\sqrt{1079}}{108}i,\left(\frac{47}{108}-\frac{7\sqrt{1079}}{116532}i,\frac{\sqrt{82239222}}{19422}\right)\right)$
&
$\leftrightarrow$
&
$[54{,}29{,}22{,}18{,}19{,}17]\teq [T_{\scalebox{0.65}{$10$}}$]
\\[3.5pt]

$\tau_{29}\!\!=$ &
$\left(\frac{5}{17}+\frac{2\sqrt{15}}{17}i,\left(\frac{47}{68}+\frac{193\sqrt{15}}{4080}i,\frac{\sqrt{3239265}}{4080}\right)\right)$
&
$\leftrightarrow$
&
$[68{,}48{,}22{,}48{,}20{,}29]\teq [T_{\scalebox{0.65}{$7$}}$]
\\[3.5pt]

$\tau_{30}\!\!=$ &
$\left(\frac{41}{88}+\frac{\sqrt{2543}}{88}i,\left(\frac{21}{44}+\frac{1251\sqrt{2543}}{111892}i,\frac{\sqrt{710682038}}{55946}\right)\right)$
&
$\leftrightarrow$
&
$[44{,}36{,}34{,}24{,}27{,}10]\teq [T_{\scalebox{0.65}{$19$}}$]
\\[3.5pt]

$\tau_{31}\!\!=$ &
$\left(\frac{57}{116}+\frac{\sqrt{2319}}{116}i,\left(\frac{15}{29}+\frac{566\sqrt{2319}}{67251}i,\frac{\sqrt{854289453}}{67251}\right)\right)$
&
$\leftrightarrow$
&
$[58{,}36{,}34{,}25{,}24{,}11]\teq [T_{\scalebox{0.65}{$16$}}$]
\\[3.5pt]

$\tau_{32}\!\!=$ &
$\left(\frac{43}{88}+\frac{\sqrt{1759}}{88}i,\left(\frac{23}{44}+\frac{859\sqrt{1759}}{77396}i,\frac{\sqrt{491580694}}{38698}\right)\right)$
&
$\leftrightarrow$
&
$[88{,}72{,}68{,}43{,}41{,}29]\teq [T_{\scalebox{0.65}{$34$}}$]
\\[3.5pt]

$\tau_{33}\!\!=$ &
$\left(\frac{20}{41}+\frac{3\sqrt{33}}{41}i,\left(\frac{43}{82}-\frac{20\sqrt{33}}{4059}i,\frac{\sqrt{17187159}}{8118}\right)\right)$
&
$\leftrightarrow$
&
$[41{,}22{,}20{,}18{,}17{,}19]\teq [T_{\scalebox{0.65}{$31$}}$]
\\[3.5pt]

$\tau_{34}\!\!=$ &
$\left(\frac{19}{72}+\frac{\sqrt{1079}}{72}i,\left(\frac{49}{72}+\frac{473\sqrt{1079}}{77688}i,\frac{\sqrt{27413074}}{12948}\right)\right)$
&
$\leftrightarrow$
&
$[72{,}54{,}20{,}48{,}22{,}29]\teq [T_{\scalebox{0.65}{$5$}}$]
\\[3.5pt]

$\tau_{35}\!\!=$ &
$\left(\frac{41}{88}+\frac{\sqrt{2543}}{88}i,\left(\frac{5}{11}+\frac{323\sqrt{2543}}{27973}i,\frac{\sqrt{710682038}}{55946}\right)\right)$
&
$\leftrightarrow$
&
$[44{,}38{,}34{,}24{,}27{,}10]\teq [T_{\scalebox{0.65}{$17$}}$]
\\[3.5pt]

\textcolor{violet}{$\tau_{36}\!\!=$} &
\textcolor{violet}{$\left(\frac{1}{2}+\frac{\sqrt{15}}{12}i,\left(\frac{55}{96}-\frac{\sqrt{15}}{240}i,\frac{\sqrt{63515}}{480}\right)\right)$}
&
\textcolor{violet}{$\rightarrow$}
&
\textcolor{violet}{$[48{,}29{,}22{,}17{,}17{,}19]\teq [T_{\scalebox{0.65}{$13$}}$]}
\\[3.5pt]

$\tau_{37}\!\!=$ &
$\left(\frac{19}{72}+\frac{\sqrt{1079}}{72}i,\left(\frac{7}{12}+\frac{101\sqrt{1079}}{12948}i,\frac{\sqrt{27413074}}{12948}\right)\right)$
&
$\leftrightarrow$
&
$[72{,}54{,}20{,}41{,}29{,}22]\teq [T_{\scalebox{0.65}{$24$}}$]
\\[3.5pt]

$\tau_{38}\!\!=$ &
$\left(\frac{21}{76}+\frac{\sqrt{1079}}{76}i,\left(\frac{45}{76}+\frac{613\sqrt{1079}}{82004}i,\frac{\sqrt{260424203}}{41002}\right)\right)$
&
$\leftrightarrow$
&
$[76{,}54{,}20{,}43{,}29{,}22]\teq [T_{\scalebox{0.65}{$26$}}]$
\\[3.5pt]

$\tau_{39}\!\!=$ &
$\left(\frac{21}{68}+\frac{\sqrt{1055}}{68}i,\left(\frac{41}{68}+\frac{669\sqrt{1055}}{71740}i,\frac{\sqrt{227828305}}{35870}\right)\right)$
&
$\leftrightarrow$
&
$[68{,}48{,}22{,}43{,}29{,}20]\teq [T_{\scalebox{0.65}{$28$}}$]
\\[3.5pt]

\bottomrule

\end{tabular}

\caption{List of similarity classes in the orbit of \([T_1]\) and their associated quintuples. The quintuples highlighted in blue and violet correspond to the same similarity class. 
}
\label{table:37}
\end{table}
\FloatBarrier

\bibliography{bibtex}

\end{document}